\documentclass[twocolumn]{aastex63}
\received{\today}
\submitjournal{The Astrophysical Journal}

\shorttitle{A cosmic Zevatron}
\shortauthors{Salamin et al.}
\watermark{Draft}
\graphicspath{{./}{figures/}}
\usepackage{bm}

\begin{document}

\title{A cosmic Zevatron based on cyclotron auto-resonance}

\correspondingauthor{Yousef I. Salamin}
\email{ysalamin@aus.edu}


\author[0000-0003-2343-4031]{Yousef I. Salamin}
\affiliation{Department of Physics, American University of Sharjah, POB 26666, Sharjah, United Arab Emirates}
\affiliation{Max-Planck-Institute for Nuclear Physics, Saupfercheckweg 1, D-69117 Heidelberg, Germany}

\author[0000-0001-7567-5350]{Meng Wen}
\affiliation{Max-Planck-Institute for Nuclear Physics, Saupfercheckweg 1, D-69117 Heidelberg, Germany}
\affiliation{Department of Physics, Hubei University, Wuhan 430062, China}
\email{meng.wen@mpi-hd.mpg.de}

\author[0000-0002-1984-1470]{Christoph H. Keitel}
\affiliation{Max-Planck-Institute for Nuclear Physics, Saupfercheckweg 1, D-69117 Heidelberg, Germany}
\email{keitel@mpi-hd.mpg.de}



\begin{abstract}

A Zevatron is an accelerator scheme envisaged to accelerate particles to ZeV energies (1 ZeV = $10^{21}$ eV). Schemes, most notably the internal shock model, have been proposed to explain the acceleration of ultra-high-energy cosmic-ray (UHECR) particles that have been sporadically detected reaching Earth since 1962. Here, the cyclotron auto-resonance acceleration (CARA) mechanism is tailored and used to demonstrate possible acceleration of particles ejected as a result of violent astrophysical processes such as the merger of a binary system or a supernova explosion. Such events result in emission of highly-energetic particles and ultra-intense beamed radiation. In the simultaneous presence of a super-strong magnetic field, the condition for cyclotron auto-resonance may be met. Thus CARA can act like a {\it booster} for particles pre-accelerated inside their progenitor by shock waves, possibly among other means. As examples, it is shown that nuclei of hydrogen, helium, and iron-56, may reach ZeV energies by CARA, under which conditions the particles, while gyrating around the lines of an ultra-strong magnetic field, also surf on the waves of a super-intense radiation field. When radiation-reaction is taken into account, it is shown that the ZeV energy gained by a particle can fall off by less than an order-of-magnitude if the resonance condition is missed by roughly less than 20\%.
\end{abstract}



\section{Introduction} \label{sec:intro}

Cosmic rays are particles which reach Earth from deep space, with energies roughly in the range $10^8-10^{20}$ eV, and sometimes beyond \citep{nagano,harari}. As a result of collisions they make with the atmosphere, showers of secondary particles and flashes of light are produced, which can be seen by detectors on Earth \citep{linsley,honda,auger,abraham,osmanov}. The low-energy cosmic-ray particles are believed to originate in active stars 
and to gain their energies from the shock waves associated with such violent events as supernova explosions \citep{bell}. Particles which arrive with energies towards the end of the above range, may be coming from active galactic nuclei (AGN) with massive black holes at their centers, or from the violent merger of neutron stars. Particles of higher energies are believed to come from extra-galactic sources \citep{allard,ebisuzaki,fang}.

Extremely rare events of ultra-high-energy cosmic rays (UHECR) began to be detected more than 50 years ago \citep{linsley} with energies greater than $10^{18}$ eV. The flux of cosmic-ray particles gets attenuated as a result of interaction with the cosmic microwave background (CMB) radiation. Cosmic-ray protons, for example, with energy above a minimum of about $4\times10^{19}$ eV, the so-called Greisen-Zatsepin-Kuz'min (GZK) limit \citep{greisen,zatsepin} cannot reach Earth. During their arduous journey through intergalactic space, they lose energy quickly by producing lighter short-lived particles via the $\Delta^+$ resonance, due to interaction with the CMB photons. Examples include
\begin{eqnarray}
    p+\gamma_{CMB} &\to& \Delta^+ \to p+\pi^0,\quad \text{and}\\
    p+\gamma_{CMB} &\to& \Delta^+ \to n+\pi^+.
\end{eqnarray}
However, detection of particles heavier than the proton with energies that violate the GZK limit, in its simple form, constitutes no fundamental contradiction \citep{abbasi1,abbasi2}. In fact, measurements by the Pierre Auger Observatory in Argentina \citep{PierreAuger} suggest that most UHECR particles are nuclei of elements heavier than the proton. 

In 1991 the first extreme-energy-cosmic-ray (EECR) particle was detected \citep{abbasi1,abbasi2} with kinetic energy exceeding $3\times10^{20}$ eV. Detection of EECRs meant that either the particles originated in places within the radius dictated by the GZK limit, but have subsequently been accelerated further outside their progenitor by some unknown mechanism, or else that the GZK limit itself had been violated. 

The source of EECRs remains a mystery today, despite the existence of models that have been proposed to explain both where the particles come from and what mechanism of acceleration is responsible for the fantastic energies they carry \citep{hillas,drury,chen,aharonian3,honda,drury2,osmanov,hofmann1,fang,Liu_2017}. This work is concerned with the energy issue, and aims to advance the scheme of cyclotron auto-resonance acceleration (CARA) as a possible resolution to it. Auto-resonance acceleration of electrons and protons, under idealized conditions, has been around for quite some time now \citep{loeb1,loeb2,sal2,sal3,sal1,sal4,sal5}.

The aim of this paper is to demonstrate that CARA should actually work under known realistic conditions, to be described below, to accelerate cosmic-ray particles to ZeV energies. Realization of this goal may then lead to the development of a more complete astrophysical model in which CARA can be suitably incorporated as a proper mechanism for cosmic-ray acceleration. The paper does not make specific claims about the existence of the non-prohibitive astrophysical conditions where CARA can explain the energy carried by EECRs. It does, however, make the assumption that those particles get pre-accelerated before they enter the region were CARA can give them another tremendous boost.

In CARA, a charged particle gains energy from an ideally coherent, linearly-polarized, radiation field monotonically (by multi-photon absorption) if injected along the common direction of its propagation and that of a quasi-uniform magnetic field, and provided an auto-resonance condition is also satisfied to a good approximation. Auto-resonance occurs when the cyclotron frequency of the particle, around the lines of the magnetic field, matches the Doppler-shifted frequency of the radiation field it senses.

The possibility of cosmic-ray acceleration by intense electromagnetic waves has been explored in the past \citep{gunn}, but CARA per se has never been employed to explain acceleration of particles in astrophysical environments, to the best of our knowledge. For it to work for accelerating nuclei to ZeV energies, the scheme requires the simultaneous presence of mega- and giga-tesla magnetic fields, and the highest-power radiation fields believed to exist in the universe. Candidates for such an environment include the polar caps of magnetars, merging neutron stars, and magnetar-powered supernova explosions \citep{price,belcz,rosswog}. Radiation fields of the needed intensity \citep{aharonian4,aharonian,hofmann2,meszaros,hofmann1,grenier2,hofmann3} may be associated with a gamma-ray burst (GRB). As an example, consider a compact object of luminosity $10^{55}$ erg/s, beaming radiation through a circle of radius 50 m, centered on either pole. The emitted radiation in this case has intensity $I_0\sim10^{44}$ W/m$^2$.

The general problem of single-particle acceleration, in the simultaneous presence of a radiation field and a uniform magnetic field, will be formulated and its solution revisited in Sec. \ref{sec:theory}. The main working equations for CARA will be obtained in Sec. \ref{sec:equations}. As examples, acceleration of single protons, and single nuclei of helium and iron, will be investigated in Sec. \ref{sec:results}. The obtained results will be discussed further in Sec. \ref{sec:discussion}, with emphasis on scenarios that may lead to deviation from the ideal conditions of resonance, radiation loss and the associated radiation-reaction effects. Finally, our conclusions will be given in Sec. \ref{sec:conc}.

\section{Theory}\label{sec:theory} 

The theoretical background of our investigations will now be outlined, with the aim of making this work self-contained, starting with a specific representation for the electromagnetic fields \citep{sal3,sal1,sal2,sal4,sal5}. Consider a point particle, of mass $M$ and charge $+Q$, injected into a region in which a uniform magnetic field of strength $B_s$ exists parallel to the direction of propagation of a plane-wave linearly-polarized radiation field. Employing a Cartesian coordinate system, the combined magnetic and radiation fields may be written, in SI units, as
\begin{eqnarray}
    \label{E}\bm{E} &=& \hat{\bm{x}}E_0 \sin\eta,\\
    \label{B}\bm{B} &=& \hat{\bm{y}}\frac{E_0}{c} \sin\eta+\hat{\bm{z}}B_s,
\end{eqnarray}
where $\hat{\bm{x}}, \hat{\bm{y}}$, and $\hat{\bm{z}}$ are unit vectors in the $x-$, $y-$, and $z-$directions, respectively, $E_0$ and $B_s$ are constants, and $\eta=\omega t-kz$ is the phase of the plane-wave radiation field, of frequency $\omega$ and wavevector $\bm{k}$ (with $\omega = ck$).

The particle's relativistic momentum and energy will be denoted by $\bm{p} = \gamma Mc\bm{\beta}$ and ${\cal E} = \gamma Mc^2$, respectively, where $\bm{\beta}$ is the velocity of the particle scaled by $c$, the speed of light in vacuum, and $\gamma = (1-\beta^2)^{-1/2}$ is the Lorentz factor. Neglecting radiation-reaction at this stage, for simplicity, the relativistic Newton-Lorentz equations (or energy-momentum transfer equations) of the particle, in the above field combination, are
\begin{eqnarray}
    \label{peq} \frac{d\bm{p}}{dt} &=& Q(\bm{E}+c\bm{\beta}\times\bm{B}),\\
\label{Eeq} \frac{d{\cal E}}{dt} &=&         Qc\bm{\beta}\cdot\bm{\bm{E}}.
\end{eqnarray}
Subject to the rather simple initial conditions of position at the origin of coordinates and injection scaled velocity $\bm{\beta} = \beta_0\hat{\bm{z}}$, these equations admit exact solutions, in closed analytic form, for the particle's trajectory and Lorentz factor. The steps leading to the desired solutions are fairly straightforward. First the $z-$ component of Eq. (\ref{peq}) and Eq. (\ref{Eeq}) read, respectively
\begin{eqnarray}
\label{pzeq} \frac{d}{dt} (\gamma\beta_z) &=& a_0\omega \beta_x\sin\eta;\quad a_0 \equiv \frac{QE_0}{M\omega c},\\
\label{eeq} \frac{d\gamma}{dt} &=& a_0\omega \beta_x\sin\eta.
\end{eqnarray}
Note that $a_0$ is a dimensionless radiation field strength, which makes $a_0^2$ a dimensionless intensity parameter. The initial conditions adopted above imply an initial value for the radiation field phase of $\eta_0=0$. With this in mind, the left-hand sides of Eqs. (\ref{pzeq}) and (\ref{eeq}) may be equated and the result integrated to yield a constant of the motion, namely
\begin{equation}\label{constant}
	\gamma(1-\beta_z) = \gamma_0(1-\beta_0); \quad \gamma_0=\frac{1}{\sqrt{1-\beta_0^2}}.
\end{equation}
The analytic solutions may best be arrived at if $\eta$ is employed to replace the time $t$ as a variable \citep{hartemann} with the following transformation playing a key role 
\begin{equation}\label{trans}
	\frac{d}{dt} = \omega(1-\beta_z)\frac{d}{d\eta}.
\end{equation}
In terms of $\eta$ and with the help of (\ref{constant}) the $x-$ and $y-$components of Eq. (\ref{peq}) now read, respectively
\begin{eqnarray}
\label{gbxe} \frac{d}{d\eta}(\gamma\beta_x) &=& a_0 \sin\eta+\alpha(\gamma\beta_y), \\
& &\nonumber\\
\label{gbye} \frac{d}{d\eta}(\gamma\beta_y) &=& -\alpha(\gamma\beta_x);\quad \alpha = \frac{Q B_s}{M\omega \gamma \left(1-\beta_z\right)}.
\end{eqnarray}
When the constant of the motion expressed by Eq. (\ref{constant}) is used, there results
\begin{equation}\label{r}
	\alpha \to r = \frac{\omega_c}{\omega}\sqrt{\frac{1+\beta_0}{1-\beta_0}}; \quad \text{and}\quad \omega_c = \frac{QB_s}{M}.
\end{equation}
Note that $\omega_c$ is the cyclotron frequency of the particle around the lines of $\bm{B}_s$, making $r$ the ratio of the cyclotron frequency of the particle to the Doppler-shifted frequency of the radiation field sensed by the particle. When Eqs. (\ref{gbxe}) and (\ref{gbye}) are solved simultaneously, subject to the same initial conditions, they yield
\begin{eqnarray}
	\label{gbx} \gamma\beta_x &=& a_0 \left[\frac{\cos \eta-\cos (r\eta )}{r^2-1}\right],\\
\label{gby} \gamma\beta_y &=& a_0\left[\frac{\sin (r\eta )-r \sin \eta}{r^2-1}\right].
\end{eqnarray}

These equations give the $x-$ and $y-$ components of the particle's momentum, scaled by $Mc$. With the help of (\ref{constant}), (\ref{trans}) and (\ref{gbx}), Eq. (\ref{pzeq}) may be integrated, subject to same set of initial conditions, to give the $z-$component of the scaled momentum
\begin{widetext}
\begin{equation}\label{gbz}
	\gamma\beta_z = \gamma_0\beta_0+a_0^2  \gamma_0(1+\beta_0)\left[\frac{\sin^2\eta}{2\left(r^2-1\right)}
	+\frac{1-r \sin\eta \sin (r\eta)-\cos\eta \cos
   (r\eta)}{\left(r^2-1\right)^2}\right].
\end{equation}
\end{widetext}

Parametric equations will next be derived which give the particle's coordinates fully analytically. An expression for the $x-$coordinate, to begin with, may be obtained with the help of the transformation
\begin{equation}\label{dxdeta}
	\frac{dx}{d\eta} = \frac{dx/dt}{d\eta/dt}=\frac{c\beta_x}{\omega(1-\beta_z)}=\frac{c}{\omega}\gamma_0(1+\beta_0)(\gamma\beta_x).
\end{equation}
Using Eq. (\ref{gbx}) in Eq. (\ref{dxdeta}) and carrying out the integration over $\eta$, gives an expression for $x(\eta)$. Expressions for $y(\eta)$ and $z(\eta)$ may also be obtained along similar lines. Finally, one gets the following parametric equations for the particle trajectory and Lorentz factor
\begin{widetext}
\begin{equation}\label{x}
	x(\eta) = \frac{ca_0}{\omega}\gamma_0(1+\beta_0) \left[\frac{r\sin\eta-\sin(r\eta)}{r(r^2-1)}\right],
\end{equation}
\begin{equation}\label{y}
	y(\eta) = \frac{ca_0}{\omega}\gamma_0(1+\beta_0) \left[\frac{1+r^2\cos\eta-\cos(r\eta)-r^2}{r(r^2-1)}\right],
\end{equation}
\begin{equation}\label{z}
	z(\eta) = \frac{c}{\omega}\left(\frac{1+\beta_0}{1-\beta_0}\right) \left\{\left[\frac{\beta_0}{1+\beta_0}
		+\frac{a_0^2}{4}\frac{3+r^2}{(r^2-1)^2}\right]\eta
		 -\frac{a_0^2}{8}\frac{\sin(2\eta)}{(r^2-1)}+a_0^2\left[\frac{(1+r^2)\cos(r\eta)\sin\eta-2r\cos\eta\sin(r\eta)}{(r^2-1)^3}\right]\right\},
\end{equation}
\begin{equation}\label{gamma}
	\gamma(\eta) = \gamma_0\left\{1+\frac{a_0^2}{2}(1+\beta_0)\left[\frac{[\cos(r\eta)-\cos\eta]^2 +[r\sin\eta-\sin(r\eta)]^2}{(r^2-1)^2}\right]\right\}.
\end{equation}
\end{widetext}

From Eqs. (\ref{x})--(\ref{gamma}) stem all aspects of the particle dynamics in the magnetic and radiation fields expressed by Eqs. (\ref{E}) and (\ref{B}). It ought to be recognized that, viewed as a function of the parameter $\eta$, the time $t=\eta/\omega+z(\eta)/c$ is a highly transcendental equation. Nevertheless, Eqs. (\ref{x})--(\ref{gamma}) can still be used to investigate evolution in time of the particle's velocity, momentum, energy, and its trajectory during interaction with the given fields. These equations will next be discussed under auto-resonance conditions.

\section{CARA: The working equations}\label{sec:equations}

Equations (\ref{x})--(\ref{gamma}) have finite limits as $r\to1$. This leads to the much anticipated auto-resonance condition $r=1$. On resonance, the solutions take on the following forms, obtained by taking the limit as $r\to1$ in Eqs. (\ref{x})--(\ref{gamma}), respectively
\begin{widetext}
\begin{equation}\label{xres}
x(\eta) = \frac{ca_0}{2\omega}\gamma_0(1+\beta_0) \left[\sin\eta-\eta\cos\eta\right],
\end{equation}
\begin{equation}
\label{yres}
y(\eta) = \frac{ca_0}{2\omega}\gamma_0(1+\beta_0) \left[\eta\sin\eta+2\cos\eta-2\right],
\end{equation}
\begin{equation}
\label{zres}
z(\eta) = \frac{c}{\omega}\left(\frac{1+\beta_0}{1-\beta_0}\right) \left\{\left(\frac{\beta_0}{1+\beta_0}\right)\eta+\left(\frac{a_0^2}{24}\right)\eta^3+\frac{a_0^2}{16}\left[2\eta\cos^2\eta-\sin(2\eta)\right]\right\},
\end{equation}
\begin{equation}
\label{gammares}
\gamma(\eta) = \gamma_0\left\{1+\frac{a_0^2}{8}(1+\beta_0)\left[\eta^2+\sin^2\eta-\eta\sin(2\eta)\right]\right\}.
\end{equation}
\end{widetext}

Equations (\ref{xres})-(\ref{gammares}) contain oscillatory as well as secular terms in the phase variable $\eta$. The purely secular term in the expression for $y$ will be responsible for a small transverse drift in the $y$-direction. A much more substantial drift in $z$ is expected due to the first and second terms in Eq. (\ref{zres}). More importantly, terms of similar nature in Eq. (\ref{gammares}) will result in tremendous energy gains by the particle.

It is very important to note that if the resonance condition is met initially, it continues to hold indefinitely. This is guaranteed by the constant of the motion expressed by Eq. (\ref{constant}) on account of the fact that $\beta\simeq\beta_z$, for ultrarelativistic particles.

An order-of-magnitude estimate for the energy gain of a nucleus, of atomic number $Z$ and mass number $A$, may be obtained, in ZeV, from Eq. (\ref{gammares}). The rapid increase in the energy gain, $\Delta{\cal E}=[\gamma(\eta)-\gamma_0] Mc^2$, is due mainly to the secular term involving $\eta^2$ in Eq. (\ref{gammares}). It can be shown that, after values of the universal constants and $\beta_0\sim1$ have been inserted, the phase-averaged energy gain, $\langle\Delta{\cal E}\rangle$, becomes 
\begin{equation}\label{Eapp}
    \langle\Delta{\cal E}\rangle[ZeV] \simeq 507\gamma_0\left(\frac{Z^2}{A}\right) \left(I_{44}\lambda^2\right) \left(1+\eta^2\right),
\end{equation} 
with $I_{44}=c\epsilon_0E_0^2/2$ the radiation field peak intensity in units of $10^{44}$ W/m$^2$, and $\lambda$ the wavelength in $\mu$m. Based on Eq. (\ref{Eapp}) a proton and a helium nucleus gain roughly the same energy after interaction with the same number of cycles of the same radiation field, but different resonance magnetic fields. This is quite evident in Figs. \ref{fig2} and \ref{fig3} as well as others.

\section{Results}\label{sec:results}

Equations (\ref{xres})--(\ref{gammares}) will now be used in accurate on-resonance single-particle calculations. Meeting the resonance condition can be a delicate matter. So, part of this section will be devoted to investigating dependence of the end results, of interest to us in this work, on fluctuations around resonance. The radiation loss and radiation-reaction effects will also be discussed.

\begin{figure}
	\centering
	\includegraphics[width=8cm]{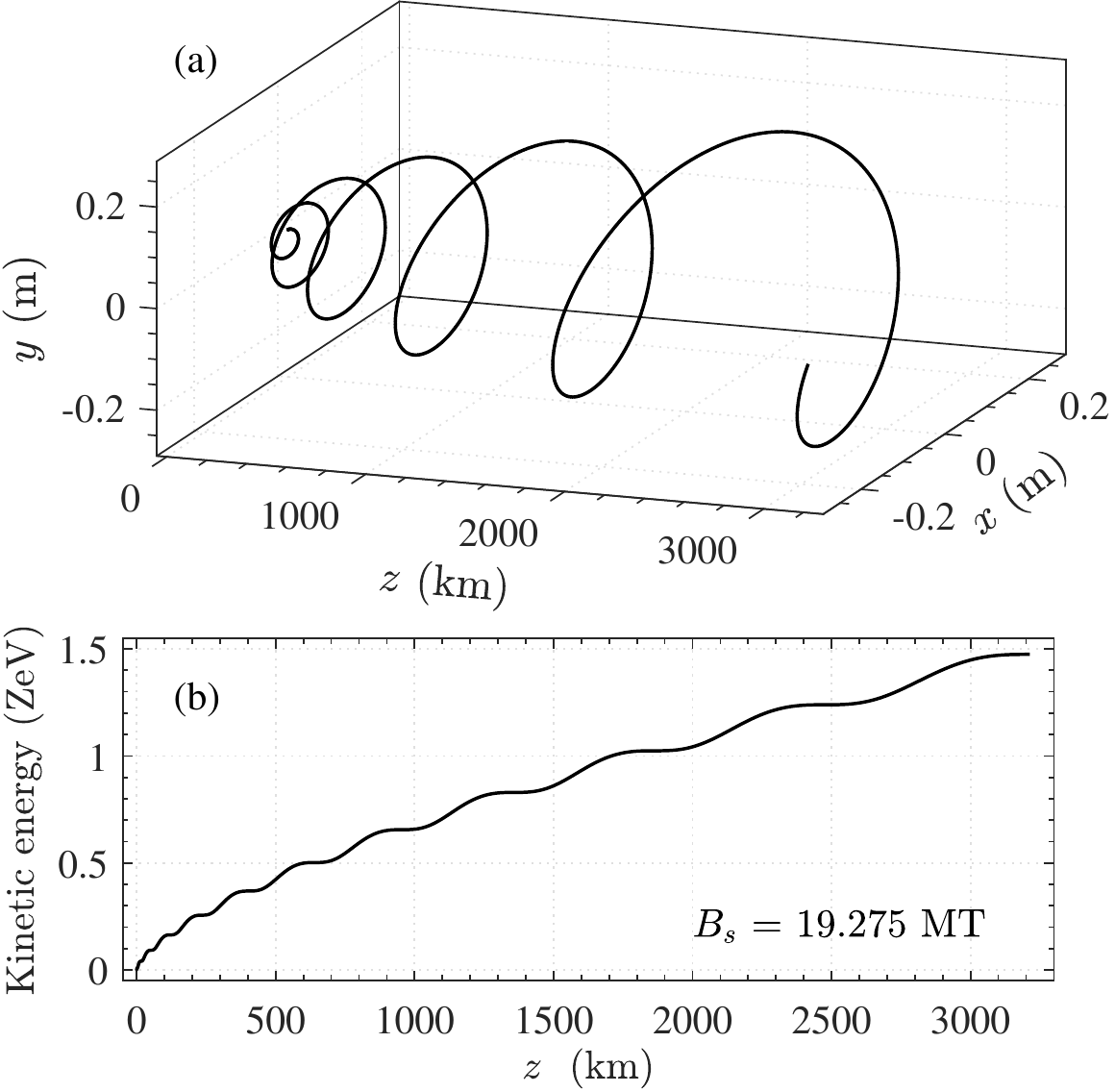}
	\caption{Example illustrating single-proton cyclotron auto-resonance acceleration (CARA). (a) Trajectory, and (b) Kinetic energy, $K = (\gamma-1)Mc^2$, in the simultaneous presence of a uniform magnetic field of strength $B_s$, and a plane-wave, linearly-polarized, radiation field of wavelength $\lambda = 1~\mu$m and intensity $I = 4\times10^{38}$ W/m$^2$. The initial proton speed is $\beta_0 = 0.02$, and evolution is followed over 6 phase cycles of the radiation field ($\Delta\eta = 12\pi$).}
	\label{fig1}
\end{figure}

\subsection{Examples}

Single-particle dynamics, in several specific magnetic and radiation field environments, will be investigated in this section, on the basis of Eqs. (\ref{xres})--(\ref{gammares}).
Typically, a particle gyrates around the lines of the magnetic field, and follows a semi-helical trajectory of increasing cross-sectional area, as Fig. \ref{fig1}(a) aims to demonstrate. For the parameters used in this example, the maximum transverse extension of the semi-helical trajectory is about 0.2 m, while the total axial excursion is over $3,200$ km. So, the particle's path is essentially a straight-line. On the other hand, the particle's kinetic energy increases monotonically, by continuous multi-photon absorption from the radiation field. For the parameters employed, a proton's kinetic energy reaches 1.5 ZeV, as shown in Fig. \ref{fig1}(b). The trajectory and evolution of the kinetic energy are shown for interaction with 6 phase $\eta$-cycles (one phase cycle $=2\pi$). The kinetic energy curve exhibits 12 {\it knees}, each of which representing a {\it kick} in the particle's energy, due to interaction with one-half of a phase cycle. Figure 1(a) also shows a semi-helical path of 6 windings. Each winding is the result of interaction with one complete phase cycle. There exists a one-to-one correspondence between the kicks in Fig. 1(b) and the windings of the semi-helical trajectory (each winding corresponds to two successive kicks).

\begin{figure}
	\centering
\includegraphics[width=8cm]{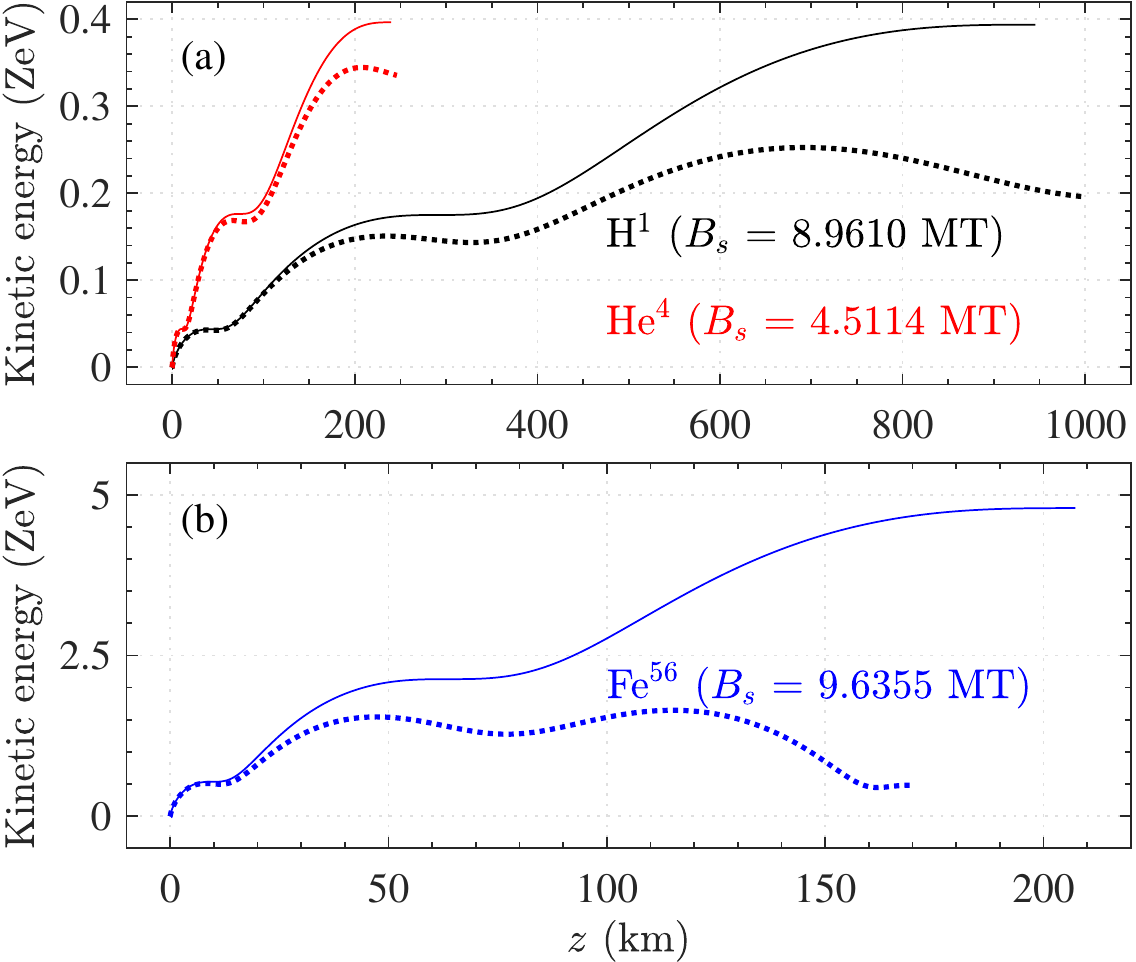}
\caption{Kinetic energy evolution with axial excursion distance. Results are shown for three nuclei, each interacting with a uniform magnetic field of strength $B_s$, and a plane-wave, linearly-polarized (visible) radiation field of wavelength $\lambda = 1~\mu$m, and intensity $I = 4\times10^{38}$ W/m$^2$. For all particles, the initial injection speed is $\beta_0 = 0.9$, and evolution is followed over 1.5 phase cycles. Every {\it knee} in each figure represents a jump in the energy gained, as a result of interaction with one-half of a phase cycle of the radiation field, or $\Delta\eta=\pi$. Solid curves: without radiation reaction, and dotted curves: with radiation-reaction, calculated using Eqs. (\ref{peq2}) and (\ref{Eeq2}).}
	\label{fig2}
\end{figure}

Figure \ref{fig1} is meant to illustrate CARA. The magnetic field strength needed to achieve auto-resonance, in this particular example, is $B_s \sim 1.92753\times10^7$ T, the likes of which are believed to be associated with {\it classical pulsars} \citep{reis:ASP01}. In Figs. \ref{fig2} and \ref{fig3}, the parameter values employed could be associated with magnetar-powered supernovae, according to recent studies \citep{greiner,sukhbold1,sukhbold}. In such environments, magnetic fields of strength $10^7-10^{11}$ T, and luminosities as high as $10^{46}$ J/s, seem to exist. Most of the energy released in such supernova explosions is radiant, carried away predominantly in the form of GRBs \citep{kumar,wang,guessoum} in addition to other radiation, including visible. Figure \ref{fig2} is for super-intense visible light, and Fig. \ref{fig3} employs parameters akin to GRBs. Note that for protons to be accelerated by CARA to ZeV energies, the required mega- or giga-tesla magnetic fields need to be present over much longer distances than is currently believed to be in existence. By comparison, acceleration of an iron nucleus takes place over a much shorter distance. More realistic examples will be presented below.

\subsection{Sensitivity of the resonance condition}

\begin{figure}
\centering
\includegraphics[width=8cm]{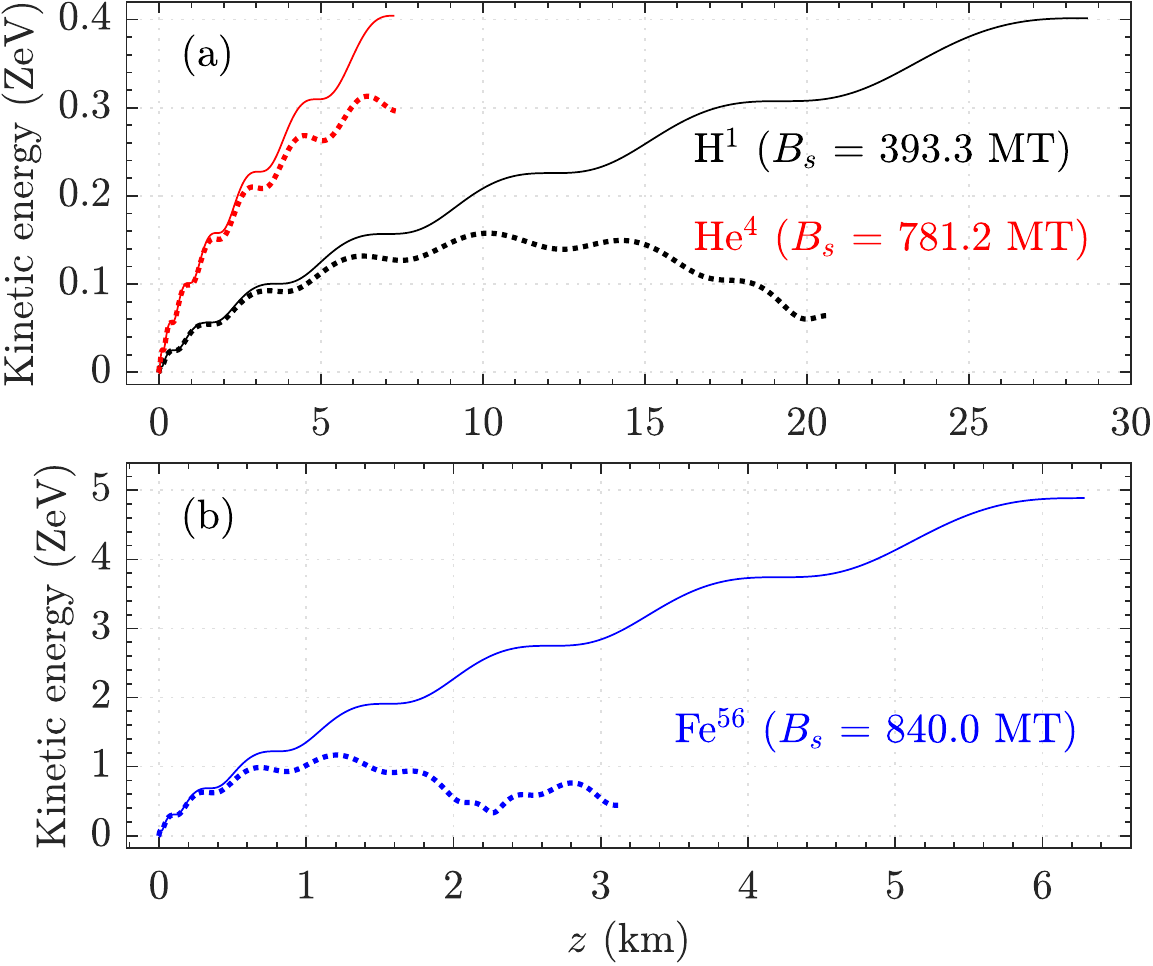}
\caption{Kinetic energy evolution with axial excursion distance. Results are for three nuclei, during interaction with a magnetic field of strength $B_s$ and a GRB of wavelength $\lambda=5\times10^{-11}$ m, and intensity $I=10^{44}$ W/m$^2$. For all particles, the injection energy $\mathcal{E}_0=500 Mc^2$, and evolution is over $\Delta\eta=8\pi$ (4 field cycles). Solid and dotted curves indicate results without and with radiation-reaction, respectively.} 
\label{fig3}
\end{figure}

Next, the seemingly sensitive dependence of the kinetic energy of a specific particle species, identified by its charge-to-mass ratio, $Q/M$, on variations, $\Delta r$, around the resonance value of $r=1$, will be investigated. Since little has been particularly specific, so far in this work, about the magnetic and radiation field environments in which the particles are accelerated, only general statements can be made about the parameter space centered about the values which lead to resonance.  As can readily be seen in Eq. (\ref{r}), the resonance condition depends on the static magnetic field $B_s$, the injection speed $\beta_0$ (or, equivalently, the scaled injection energy $\gamma_0$), and the radiation frequency $\omega$ (or, equivalently, the wavelength $\lambda$). Holding $Q/M$ fixed, Eq. (\ref{r}) yields
\begin{equation}\label{dev}
	\frac{\Delta r}{r} = \frac{\Delta B_s}{B_s}+\frac{\Delta \omega}{\omega}+\gamma_0^2 \Delta\beta_0,
\end{equation}
where the symbol $\Delta X$ stands for a spread in the possible values of the parameter $X\in\{B_s, \omega, \beta_0\}$ around its on-resonance value\footnote{We note that the resonance condition may be violated by other effects such as particle-particle collisions, as well as lack of coherence and other deviations from our assumptions about the radiation field. While those issues need to be investigated in future publications, our investigations in the present work, based on the parameter set $\{B_s, \omega, \beta_0\}$, already hint clearly that small deviations from resonance are not very critical for explaining cosmic ZeV acceleration by CARA.}. Thus, variations in any one, or more, of the parameters in this set, lead to a departure from $r=1$. This means that investigation of the dynamics over a set of values of $r$ around unity is tantamount to studying the dynamics over many sets of different values of these quantities, simultaneously. As a first example, variations in the magnetic field only will be considered. Keeping $\beta_0$ and $\omega$ fixed, values of $r$ are varied for the three nuclear species  H$^{+1}$, He$^{+2}$ and Fe$^{+26}$. If one lets $r\to 1+\Delta r$ in Eqs. (\ref{x})--(\ref{gamma}) the particle dynamics may be investigated over values of $r\in[1-\Delta r,1+\Delta r]$, for the given charge-to-mass ratios.

\begin{figure}
\centering
\includegraphics[width=8cm]{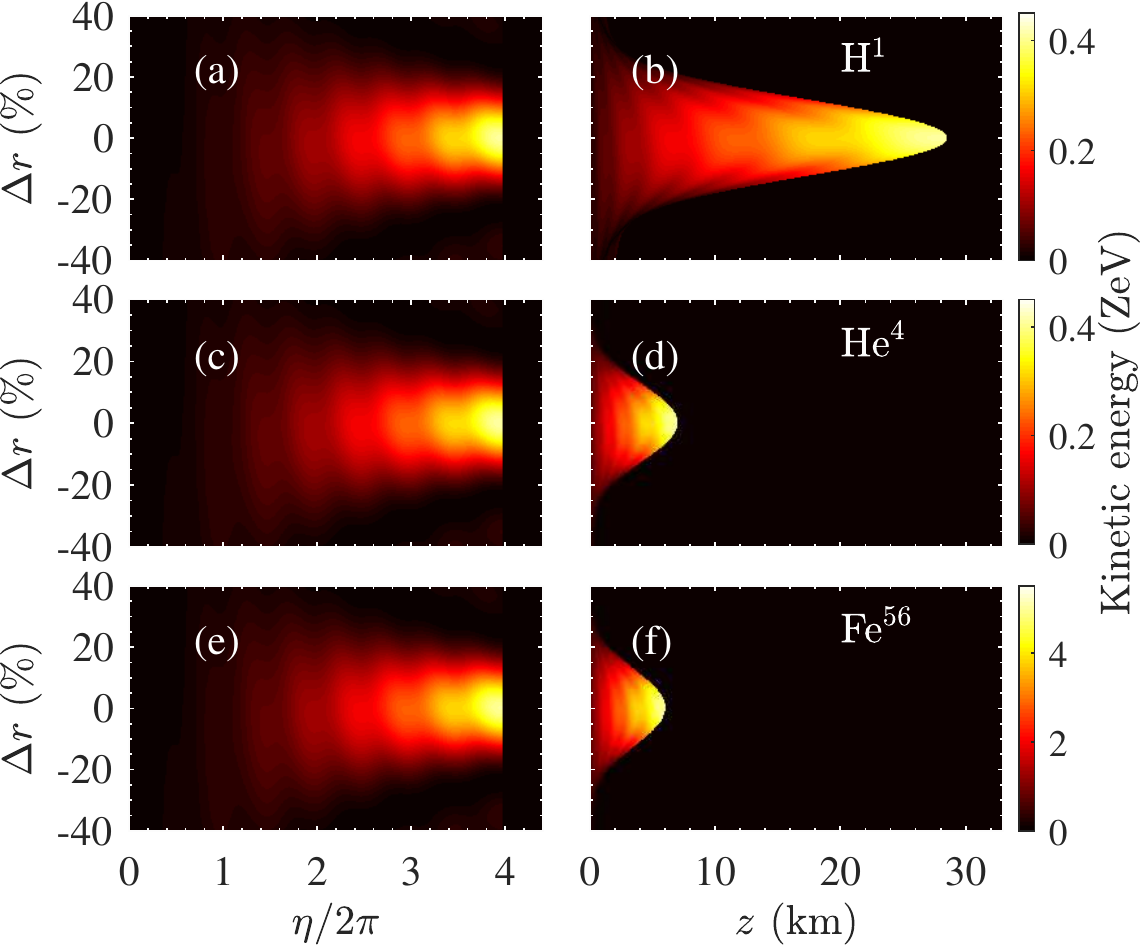}
\caption{Density plots of the kinetic energy evolution, in the absence of radiation-reaction, with time and with the axial excursion distance, both as values of $r$ are tuned around resonance by $\Delta r\in[-0.4,0.4]$, resulting from variations in the values of $B_s$, while those of $\omega$ and $\beta_0$ are kept fixed. Here, $\Delta r/r=\Delta B_s/B_s$, according to Eq. (\ref{dev}). The displayed results are for the nuclei H$^{+1}$, He$^{+2}$ and Fe$^{+26}$. Each particle interaction is with a magnetic field of strength $B_s$ centered around its corresponding on-resonance value (see Fig. \ref{fig3}) and a GRB of wavelength $\lambda=5\times10^{-11}$ m and intensity $I=10^{44}$ W/m$^2$. For all particles, the injection energy $\mathcal{E}=500Mc^2$, and evolution is over $\Delta\eta=8\pi$ (4 phase cycles). For each particle species, the spread in the values of $B_s$ is responsible for the detuning $\Delta r$. So, the displayed results cover a substantial distribution of magnetic field values, as may be expected near the polar cap of a compact object like a neutron star.}
\label{fig4}
\end{figure}

\begin{figure}
\centering
\includegraphics[width=8cm]{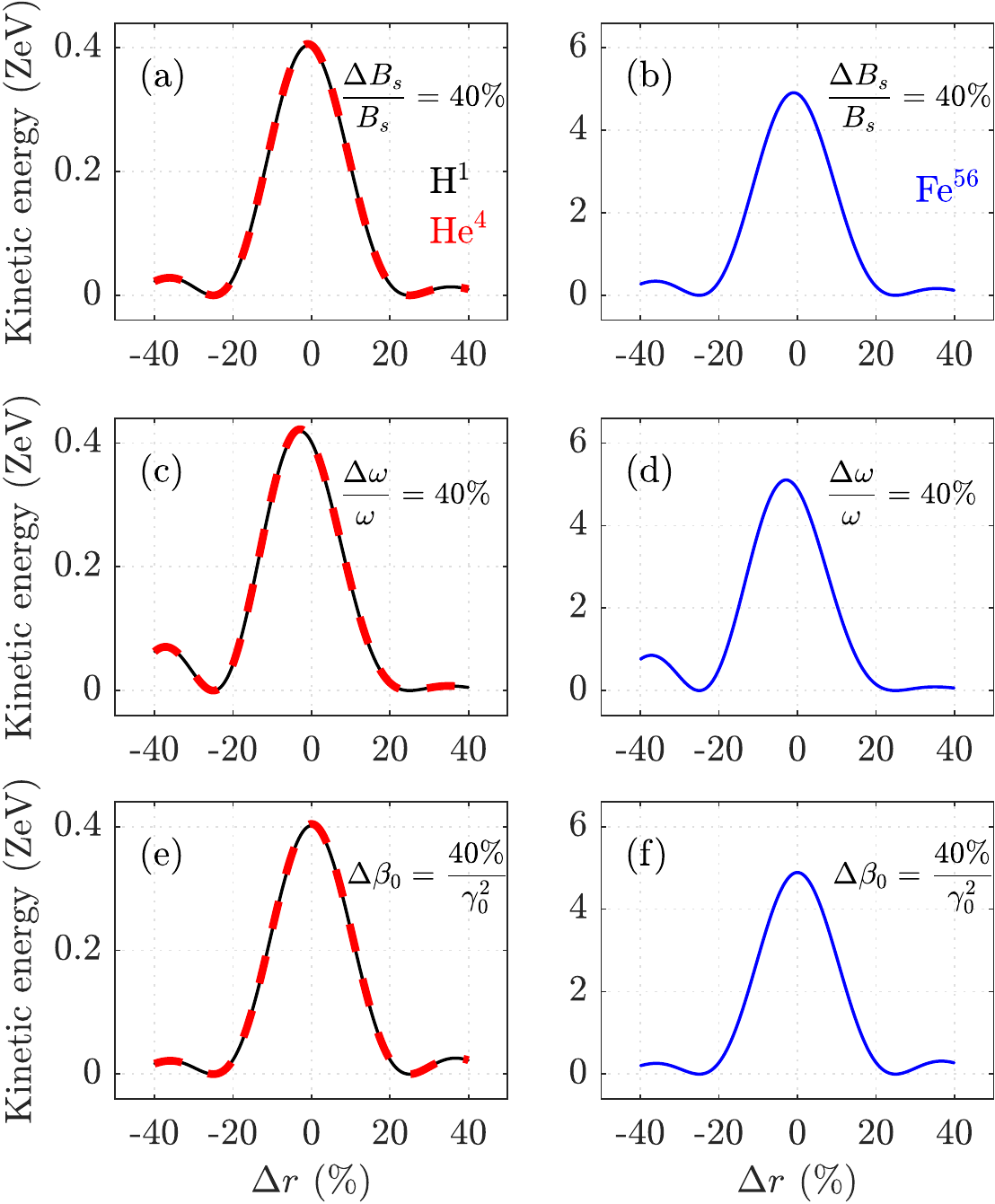}
\caption{The accelerated particle kinetic energy (in the absence of radiation-reaction) at the end of interaction with 4 cycles of the radiation field, is given as a function of the detuning $\Delta r$. Left panel is for each of the nuclei of H$^{+1}$, He$^{+2}$, and the panel on the right is for the nucleus Fe$^{+26}$. Rows 1--3 are, respectively, for detuning $\Delta r/r=\Delta B_s/B_s$, $\Delta r/r=\Delta\omega/\omega$, and $\Delta r/r=\gamma_0^2\Delta\beta_0$. The corresponding excursion distance, in each case, may be read off the corresponding plot in Fig. \ref{fig3}. All of the remaining parameters used here are the same as in Fig. \ref{fig4}.} 
	\label{fig5}
\end{figure}

In Fig. \ref{fig4}, variations in the particle's {\it exit} kinetic energy are shown as a function of the detuning $\Delta r$ around resonance ($r=1$) with the interaction time (expressed indirectly through $\eta$) and also with the distance over which the particle interaction with the magnetic and radiation fields takes place. Fig. \ref{fig5} shows the kinetic energy of each particle, at the end of its interaction with the given on-resonance magnetic field and 4 phase cycles, as a function of the detuning $\Delta r$.

Note that Eq. (\ref{r}) implies that $\Delta r/r=\Delta B_s/B_s$, for fixed values of $\omega$ and $\beta_0$. Using the magnetic field values that correspond to resonance, shown in Fig. \ref{fig3}, leads to the conclusion that a 40\% deviation, $\Delta B_s$, from its corresponding resonance values, leads to a similar percentage deviation from resonance (detuning $\Delta r$). According to Eq. (\ref{dev}) similar conclusions may be arrived at when variations, $\Delta\omega$, around the resonance value of $\omega$, are considered, keeping $B_s$ and $\beta_0$ fixed at their respective values on resonance. Finally, if one keeps $B_s$ and $\omega$ fixed at their respective on-resonance values and varies those of $\beta_0$ in the interval $[-0.4/\gamma_0^2, 0.4/\gamma_0^2]$, results in $\Delta r\sim40\%$, according to Fig. \ref{fig5}. Numerical calculations based on Eq. (\ref{gamma}) indicate that the kinetic energy at the end of the interaction, in all cases considered, remains roughly within one order-of-magnitude of its on-resonance maximum, as the values of $r$ are detuned by about 10\% $-$ 20\% on either side of $r=1$. Off-resonance conditions cause the particle to phase-slip behind the radiation wave and gain little energy from it, if at all. 

\subsection{Radiation loss}

It was remarked, introducing Fig. \ref{fig1}(a), that the trajectory of a particle accelerated by the CARA scheme can essentially be considered linear. It is also well known that, in linear accelerators on Earth, radiation loss by the accelerated particle is negligibly small, in general. The instantaneous power radiated by the particle during acceleration is given by the relativistic generalization of the Larmor formula \citep{jackson}
\begin{equation}\label{P}
    P = \frac{Q^2\gamma^6}{6\pi\epsilon_0c}\left[\left(\frac{d\bm{\beta}}{dt}\right)^2-\left(\bm{\beta}\times\frac{d\bm{\beta}}{dt}\right)^2\right].
\end{equation}

\begin{figure}
\centering
\includegraphics[width=8cm]{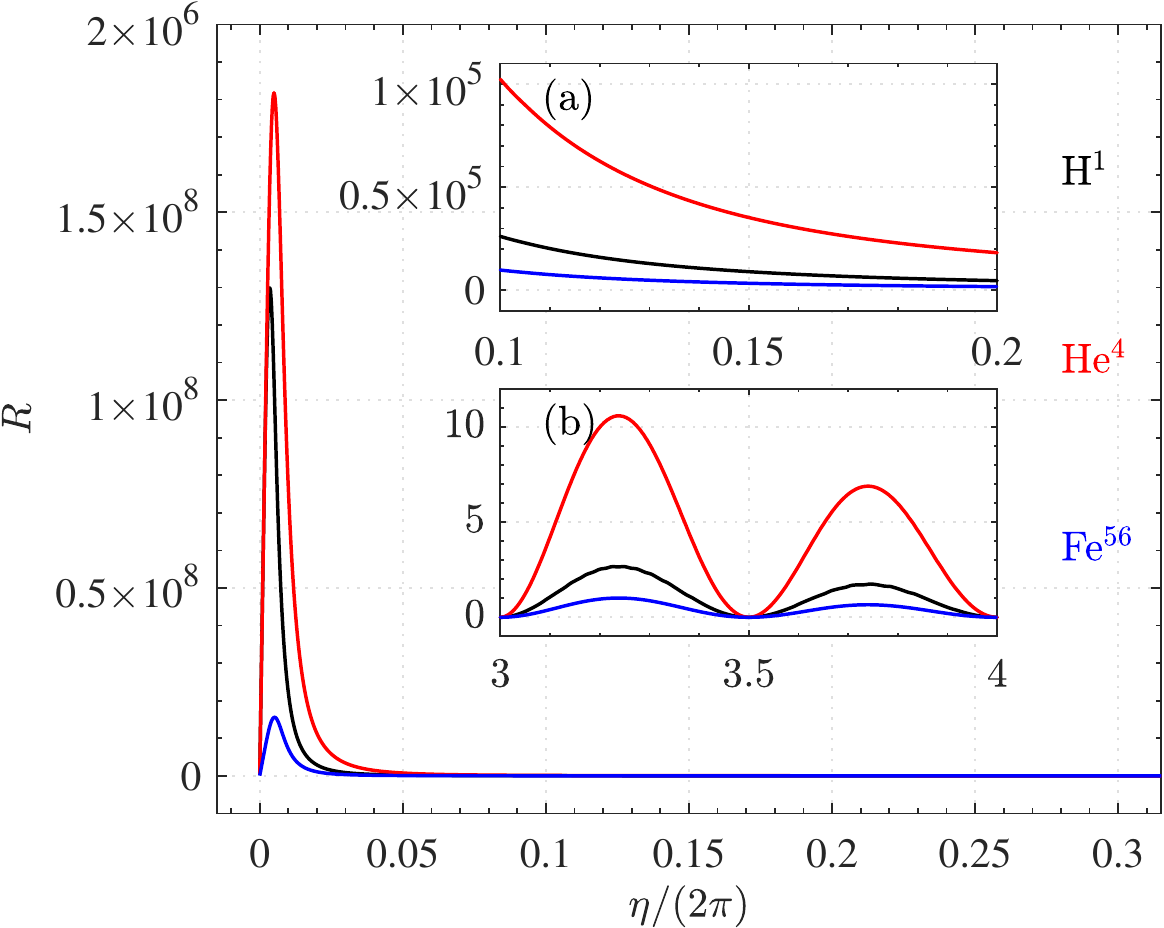}
\caption{Ratio of the rate of energy gain to the power lost by the particles during acceleration, calculated using Eq. (\ref{R}), as a function of the number of (phase) $\eta$-cycles. The results shown here are for the three nuclei H$^{+1}$, He$^{+2}$ and Fe$^{+26}$, considered in Figs. \ref{fig3}--\ref{fig5}, and employing a parameter set the same as that of Fig. \ref{fig3} (without radiation-reaction).}
\label{fig6}
\end{figure}

Presence of the factor $\gamma^{6}$ upfront in this expression can make the radiated power quite large, even exceeding the rate of energy gain, $d{\cal E}/dt$, beyond some point. To avoid running into numerical instabilities, the following alternative expression for the power, in terms of derivatives with respect to the phase $\eta$ and the other on-resonance parameters, will be used \citep{sal1}  
\begin{equation}\label{PP}
    P(\eta) = \frac{(Q\omega_c\gamma)^2}{6\pi\epsilon_0c}\left[\left(\frac{d\bm{\beta}}{d\eta}\right)^2+\gamma^2\left(\bm{\beta}\cdot\frac{d\bm{\beta}}{d\eta}\right)^2\right].
\end{equation}
Power lost through radiation will be assessed, in relation to the rate at which energy is gained by the particle, by investigating the dimensionless ratio 
\begin{equation}\label{R}
    R(\eta) = \frac{1}{P} \frac{d{\cal E}}{dt}.
\end{equation}
This ratio is shown as a function of the phase variable $\eta$ in Fig. \ref{fig6}, for the cases considered in Figs. \ref{fig3}--\ref{fig5}. According to Fig. \ref{fig6}, $R\gg 1$ during interaction with a small fraction of the first phase cycle. The corresponding interaction time over which the spike in $R$ happens is less than 10 ps, according to the inset in Fig. \ref{fig8}, to be introduced shortly below. This happens while the initial part of the particle's trajectory is still almost straight, implying a relatively small radiation energy loss rate. Subsequent to this brief spike in the values taken up by this ratio, $R$ drops sharply and may approach unity (or less) only over brief intervals during the particle's motion, as may be inferred from the insets in Fig. \ref{fig6}.

\subsection{Radiation-reaction effects}

\begin{figure}
\centering
\includegraphics[width=8cm]{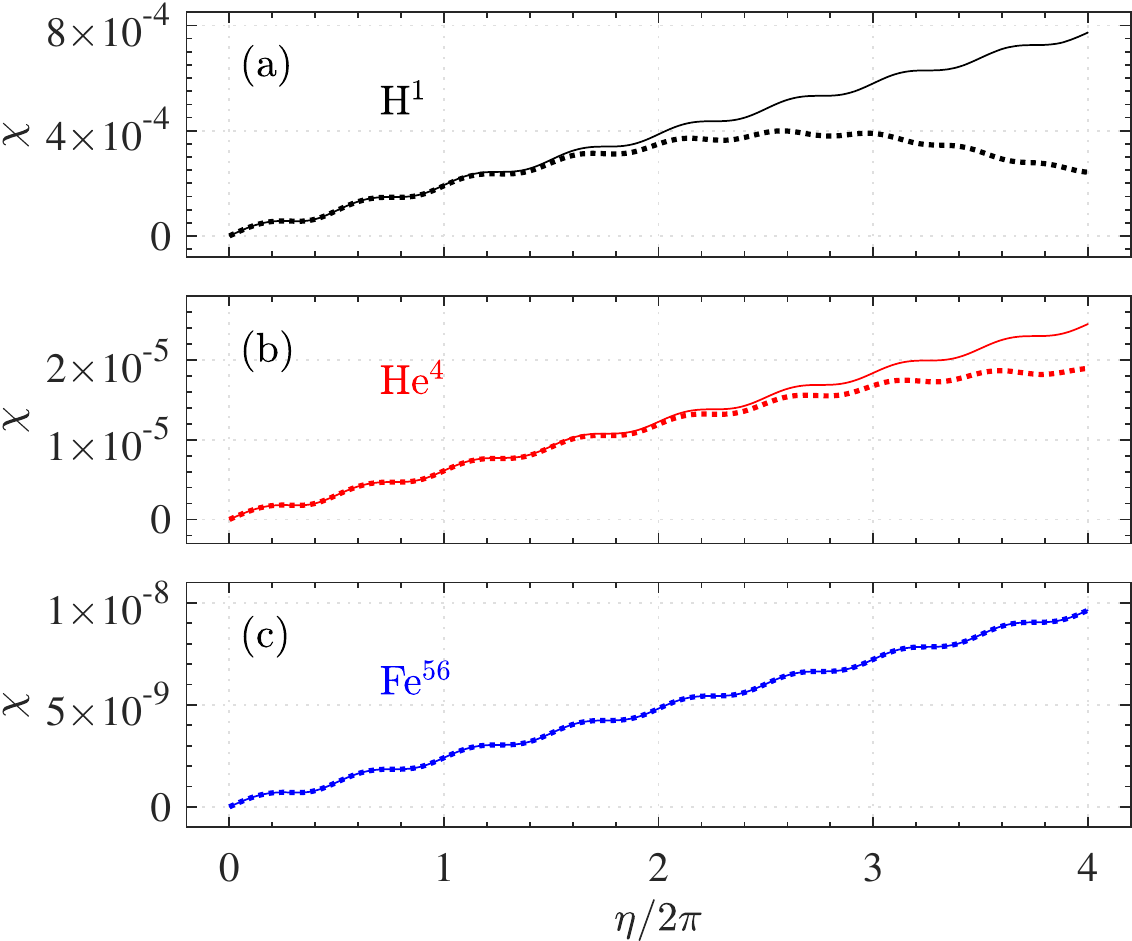}
\caption{The quantum efficiency parameter $\chi$ as a function of $\eta$ for the particles and parameter set of Fig. \ref{fig3}. The solid(dotted) curves are for without(with) radiation-reaction, respectively. For Fe$^{56}$, the two curves are indistinguishable.}
\label{fig7}
\end{figure}

At the time of emission of the radiation, the particle recoils as it experiences a radiation-reaction (RR) force, which must be taken into account as, especially in circumstances for which $R\to1$ \citep{hadad,Tamburini_2010,TAMBURINI2011,seto,tursunov,antonino}. To assess the RR effects quantitatively, Eqs. (\ref{peq}) and (\ref{Eeq}) will now be replaced by \citep{hadad,seto}
\begin{eqnarray}
\label{peq2} \frac{d\bm{p}}{dt} &=& Q(\bm{E}+c\bm{\beta}\times\bm{B})-\bm{f}_1+\bm{f}_2+\bm{f}_3,\\
\label{Eeq2} \frac{d{\cal E}}{dt} &=& Qc\bm{\beta}\cdot\bm{\bm{E}}-P_1+P_2+P_3,
\end{eqnarray}
respectively. In these equations, $\bm{f}_i$ is a force term and $P_i$ is a power term. With $i = 1, 2$ and 3, the force and power terms are given by
\begin{eqnarray}
\label{f1}\bm{f}_1 &=& \frac{\bm{\beta}}{c}P_1,\\ 
\bm{f}_2 &=& \gamma A\left[\frac{Q}{Mc^2}\right] \left[\frac{\partial}{\partial t}+c\bm{\beta}\cdot \bm{\nabla}\right]\left(\bm{E}+c\bm{\beta}\times\bm{B}\right),\\
\bm{f}_3 &=& A\left[\frac{Q}{Mc}\right]^2 \left[\left(\bm{\beta}\cdot\bm{E}\right)\frac{\bm{E}}{c}+\left(\bm{E}+c\bm{\beta}\times\bm{B}\right)\times\bm{B}\right],\\
P_1 &=& \gamma^2A  \left[\left(\frac{d\bm{\beta}}{d\eta}\right)^2+\gamma^2\left(\bm{\beta}\cdot\frac{d\bm{\beta}}{d\eta}\right)^2\right]\omega_c^2 = P(\eta), \\
P_2 &=& \gamma A\left[\frac{Q}{Mc^2}\right] \left[\frac{\partial}{\partial t}+c\bm{\beta}\cdot \bm{\nabla}\right]\left(c\bm{\beta}\cdot\bm{E}\right),\\
P_3 &=& A\left[\frac{Q}{Mc}\right]^2\left(\bm{E}+c\bm{\beta}\times\bm{B}\right)\cdot\bm{E};\quad A = \frac{Q^2}{6\pi\epsilon_0c}.
\end{eqnarray}

\begin{figure}
\centering
\includegraphics[width=8cm]{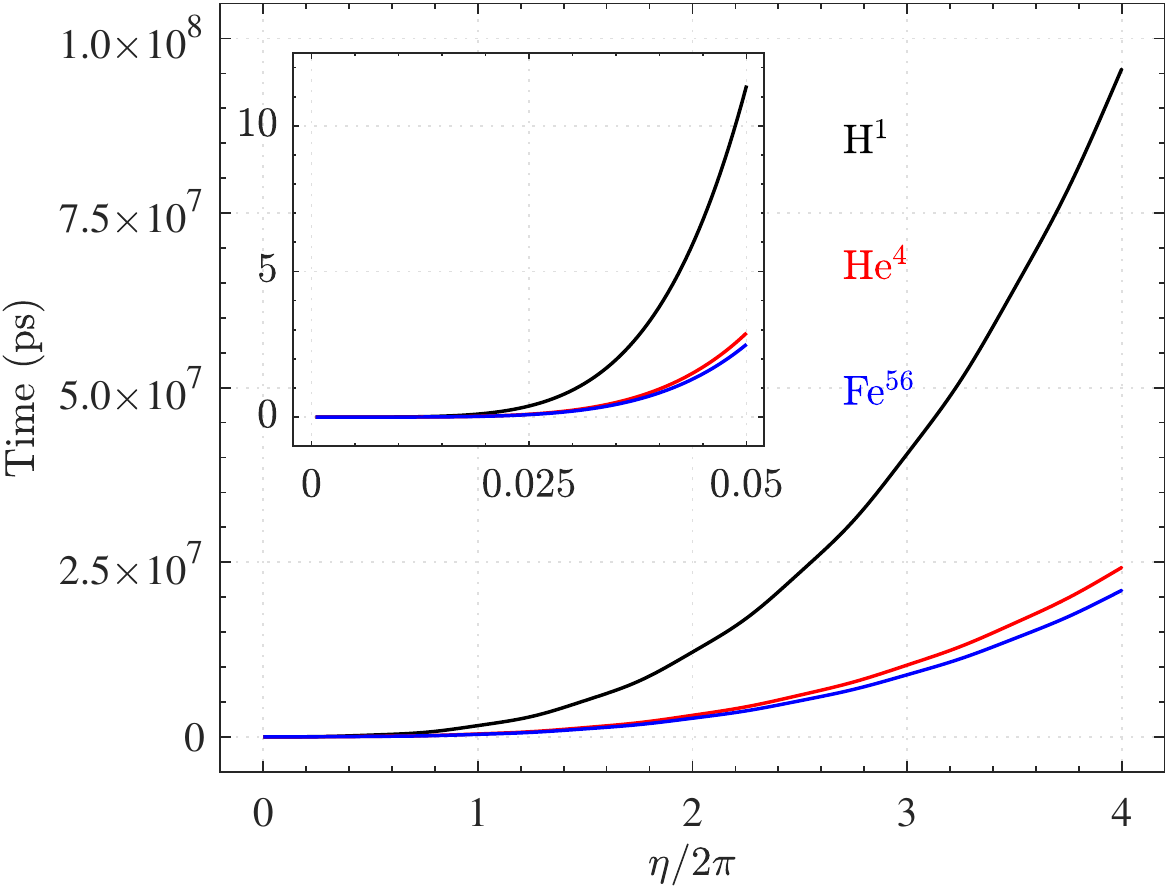}
\caption{Interaction time as a function of the number of $\eta-$cycles for the particles and parameter set of Fig. \ref{fig3} (without radiation-reaction) based on the transcendental equation $t=\eta/\omega+z(\eta)/c$.}
\label{fig8}
\end{figure}

This classical treatment of RR is assumed to be a good approximation, in light of the fact that $\chi$, the quantum efficiency parameter, can be quite small \citep{hadad, Tamburini_2010}. This parameter is defined by \citep{Ritus1985}
\begin{equation}
    \chi=\frac{Q\hbar\gamma}{M^2c^3} \sqrt{\left(\bm{E}+c\bm{\beta}\times\bm{B}\right)^2-\left(\bm{\beta}\cdot\bm{E}\right)^2},
\end{equation}
where $\hbar$ is the reduced Planck constant. For an electron of mass $M=m_e$, the RR effects can be quite sizeable. But for protons and heavier nuclei (mass $M\gtrsim2000 m_e$) of the same $\gamma-$factor, $\chi$ is at least six orders of magnitude smaller. For all of the examples considered, and for the parameters used, $\chi\ll1$, as Fig. \ref{fig7} clearly demonstrates.

Note that $\bm{f}_2$ and $P_2$ are a factor of $\gamma\sim10^{12}$ times smaller than $\bm{f}_1$ and $P_1$, respectively, for ZeV particles. Furthermore, $\bm{f}_3$ and $P_3$ are a factor of $\gamma^2\sim10^{24}$ times smaller than $\bm{f}_1$ and $P_1$, respectively. Consequently, only $\bm{f}_1$ and $P_1$ will be retained in Eqs. (\ref{peq2}) and (\ref{Eeq2}) which will subsequently be solved numerically. For ultrarelativistic particle energies and small field gradients, the dropped terms contribute negligibly. Results for the exit kinetic energies, stemming from those solutions, are displayed with dotted lines in Fig. \ref{fig2}. Effect of RR on the radiated power and energy gain by the nuclei of H$^1$, He$^4$ and Fe$^{56}$ is to lower those quantities by roughly less than one order-of-magnitude. Another notable related effect is shortening of the total excursion distance, in most cases considered, due to RR. This is reminiscent of the effect of friction on the motion of macroscopic objects.

Similar plots are displayed in Fig. \ref{fig3}. With the RR effects taken into account, all three particles are shown to leave the loosely-defined interaction region with kinetic energies in excess of $10^{20}$ eV. According to Fig. \ref{fig3}, the RR effects are responsible for a reduction in the exit kinetic energy. Reduction in the particle's excursion distance, during the same interaction cycle and with RR taken into account, is also clear in Figs. \ref{fig3}(a) and (b).

Up to this point, time dependence has been expressed indirectly in terms of the dependence upon the phase $\eta$. For the examples of Fig. \ref{fig3}, variation of the time $t$ with $\eta$ is shown in Fig. \ref{fig8}, according to which the interaction times of the nuclei of H$^1$, He$^4$ and Fe$^{56}$ are, roughly, 96 $\mu$s, 24 $\mu$s and 20 $\mu$s, respectively. In terms of the period $\tau=\lambda/c$ of the GRB employed, these are equivalent to 5.76$\times10^{14}\tau$, 1.44$\times10^{14}\tau$ and 1.2$\times10^{14}\tau$, respectively.

\begin{deluxetable}{lcccccccccccc}
\tablenum{1}
\label{tab:table1}
\tablecaption{Energy gain drop due to radiation-reaction}  
\tablewidth{0pt}
\tablehead{
\colhead{} &~&~&~& \colhead{$K$} &~&~&~& \colhead{$K'$} &~&~&~& \colhead{$\frac{\Delta K}{K}$}\\
Particle &~&~&~& (ZeV) &~&~&~& (ZeV) &~&~&~& (\%)}
\startdata
H$^{+1}$ &~&~&~& 0.4015 &~&~&~& 0.2486 &~&~&~& 38 \\
He$^{+2}$ &~&~&~& 0.4043 &~&~&~& 0.3549 &~&~&~& 12 \\
Fe$^{+26}$ &~&~&~& 4.8878 &~&~&~& 1.9941 &~&~&~& 59 \\
\enddata
\end{deluxetable}

It cannot be concluded with certainty, based on Figs. \ref{fig2} and \ref{fig3}, that the RR effects act to universally lower the accelerated particle kinetic energy so drastically in CARA. With RR taken into account according to the approximate version of Eqs. (\ref{peq2})--(\ref{Eeq2}) the right-hand sides of (\ref{pzeq}) and (\ref{eeq}) are no longer the same. Subsequently, subtraction of the RR-based equivalent to (\ref{pzeq}) from the approximate RR-equivalent to (\ref{eeq}) does not lead to a constant of the motion analogous to that expressed by Eq. (\ref{constant}) but to
\begin{equation}
\label{gammabetaz}
	\frac{d}{dt}\gamma \left(1-\beta_z\right) = - \frac{(1-\beta_z)P(t)}{Mc^2},
\end{equation}
instead. Making the $t\to\eta$ transformation in Eq. (\ref{gammabetaz}) and integrating the result formally over $\eta$ leads to
\begin{equation}\label{rr}
    \gamma(1-\beta_z)=\gamma_0(1-\beta_0)-\frac{J}{\omega Mc^2},
\end{equation}
in which 
\begin{equation}
    J(\eta)=\int_0^\eta P(\eta')d\eta'. 
\end{equation}
Using Eq. (\ref{rr}) in the second of Eqs. (\ref{gbye}), with the replacement $B_s=B_s^{res}+\Delta B_s$, where $B_s^{res}$ is the on-resonance value of $B_s$ when RR is neglected, results in

\begin{eqnarray}\label{alpha}
    \alpha = \frac{Q(B^{res}_s+\Delta B_s)}{M\omega[\gamma_0(1-\beta_0)-J/(\omega Mc^2)]} &<& \frac{Q(B^{res}_s+\Delta B_s)}{M\omega\gamma_0(1-\beta_0)}\nonumber\\
     &=&  r+\frac{Q\Delta B_s}{M\omega_D},
\end{eqnarray}
where $\omega_D$ is the Doppler-shifted frequency of the radiation wave.  

\begin{figure}
\centering
\includegraphics[width=8cm]{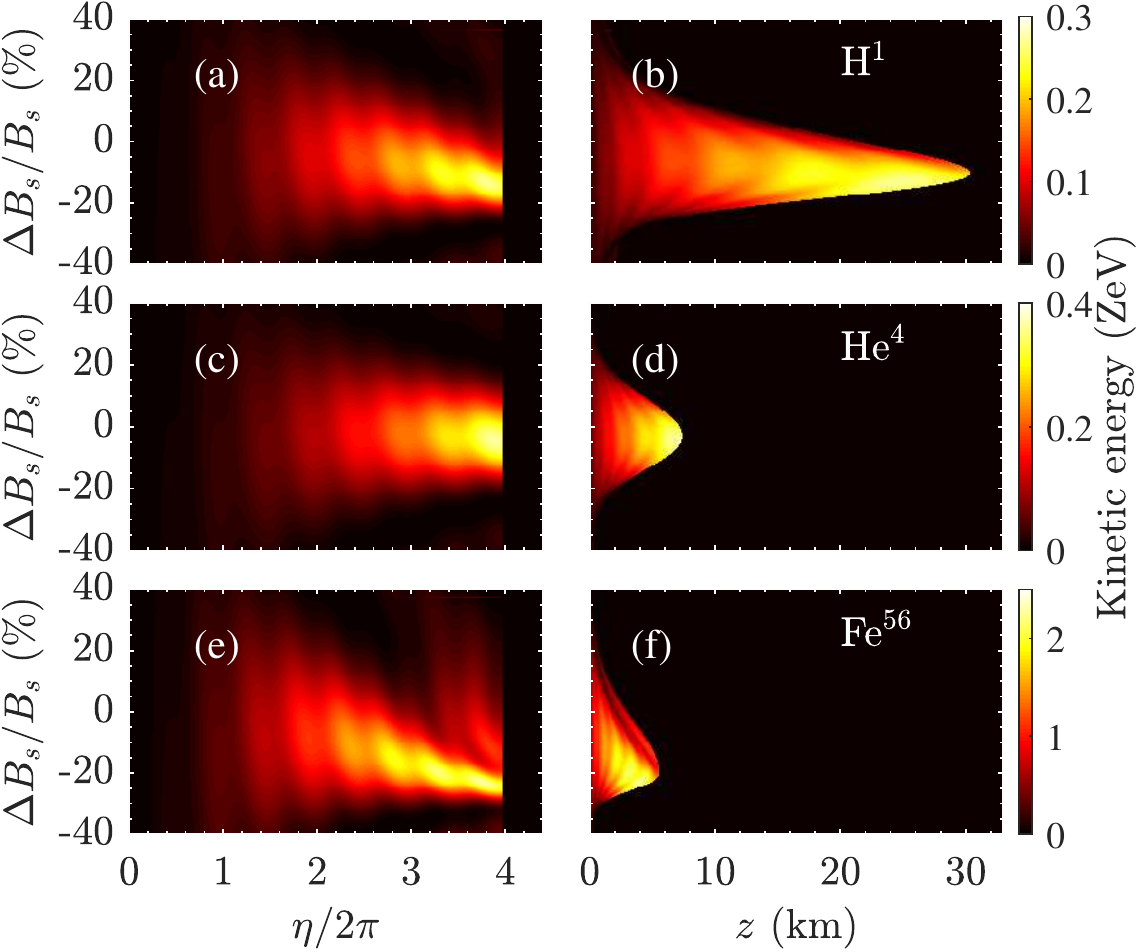}
\caption{Density plots of the kinetic energy evolution (in the presence of radiation-reaction) with time and with the axial excursion distance, both as values of the magnetic field are tuned around the on-resonance values of Fig. \ref{fig3} (without radiation-reaction) and covering $\Delta B_s/B_s\in[-0.4,0.4]$, while those of $\omega$ and $\beta_0$ are kept fixed. Each particle interaction is with a GRB of wavelength $\lambda=5\times10^{-11}$ m and intensity $I=10^{44}$ W/m$^2$. For all particles, the injection energy $\mathcal{E}=500Mc^2$, and evolution is over $\Delta\eta=8\pi$ (4 phase cycles). In principle, this figure may be compared and contrasted with Fig. \ref{fig4}, with the understanding that $\Delta B_s/B_s\ne \Delta r/r$, due to the fact that radiation-reaction is taken into account here.}
\label{fig9}
\end{figure}

\begin{figure*}
\centering
\includegraphics[width=16cm]{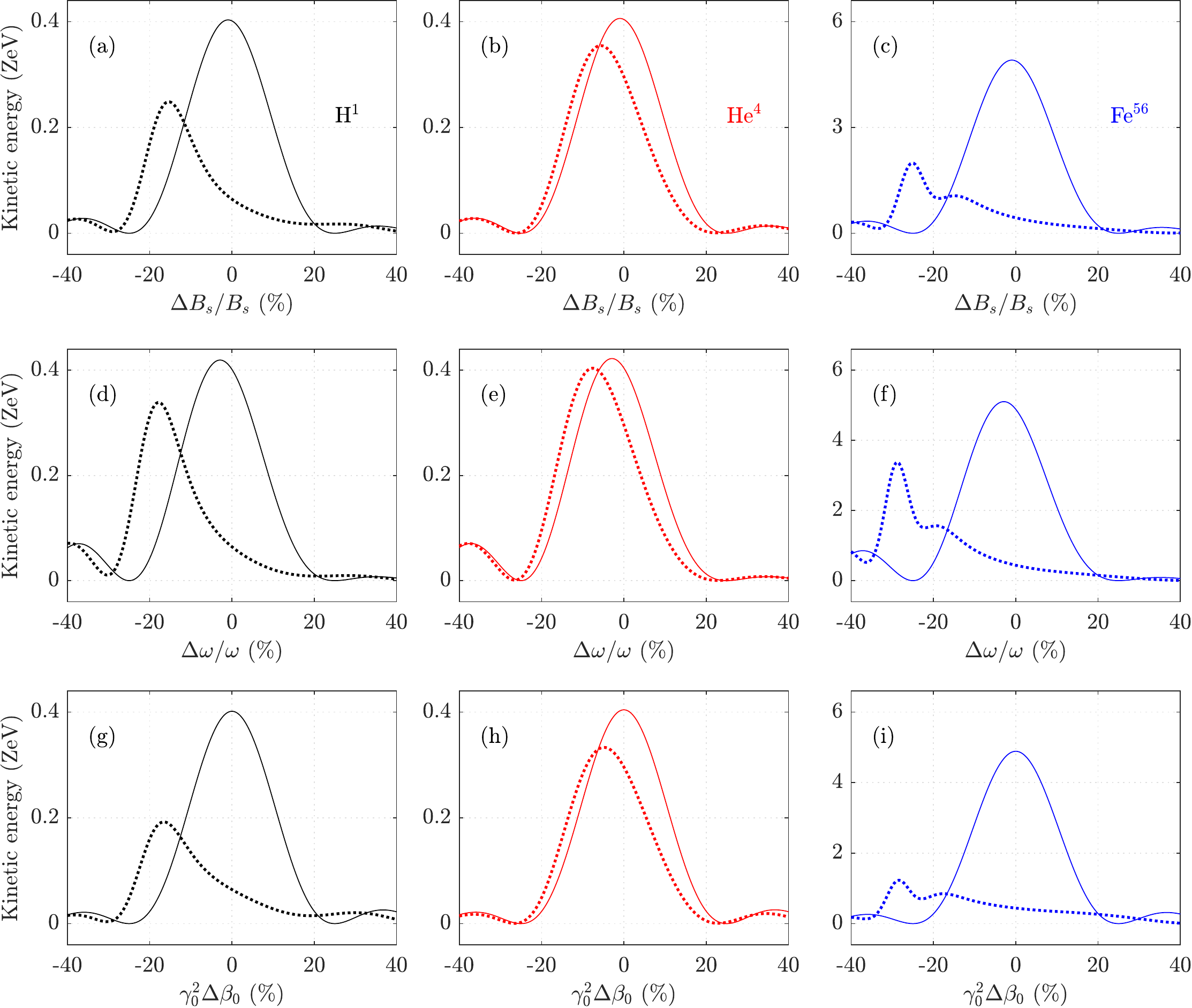}
\caption{The radiation-reaction (RR) effects shift the conditions of reaching optimal energies, from resonance expressed by $r=1$, which holds when the RR effects are neglected. Shown here is variation of the end-of-interaction particle kinetic energy which results, in each case, from a deviation, $\Delta X$, from the corresponding resonance value, determined by $r=1$, of the parameter $X\in\{B_s, \beta_0, \omega\}$. The central values, like in Fig. \ref{fig3}, are: $\omega=2\pi c/\lambda$ with $\lambda= 5\times 10^{-11}$ m, $\gamma_0^2\beta_0 = \gamma_0\sqrt{\gamma_0^2-1}$, with $\gamma_0=500$, and $B_s= 393.3$ MT, $781.2$ MT, and 840 MT,   for nuclei of H$^{+1}$, He$^{+2}$ and Fe$^{+26}$, respectively. Solid lines: without RR (as in Figs. \ref{fig4} and \ref{fig5}) and dotted lines: with RR (based on Fig. \ref{fig9}).}
\label{fig10}
\end{figure*}

In the analytic discussions of CARA in the absence of RR, $r$ plays the role of a resonance parameter. By analogy, that role would most probably be played, albeit roughly, by the quantity $\alpha$ defined by Eq. (\ref{alpha}) when RR is taken into account. Ideally, this should emerge from analytic solutions to Eqs. (\ref{peq2})--(\ref{Eeq2}). In the absence of such solutions, $\alpha$ may only be used to hint at the existence of a resonance condition different from $r=1$. 

Having been derived from the resonance condition in the absence of RR, Eq. (\ref{dev}) cannot be relied upon anymore in the presence of RR. In the subsequent analysis of the acceleration process with RR, total dependence upon the numerical solutions outlined above is inevitable.

To further assess departure from the $r=1$ resonance due to inclusion of radiation-reaction, Fig. \ref{fig9} has been produced by scanning the static magnetic field values over $\Delta B_s/B_s\in[-0.4,0.4]$ centered around the on-resonance values employed in Fig. \ref{fig3}. This figure may be read in conjunction with Fig. \ref{fig4}, but with care. In Fig. \ref{fig4}, $\Delta B_s/B_s=\Delta r/r$, which does not hold when RR is taken into account. In the absence of RR, maximum ZeV energy is reached for parameters which make the detuning $\Delta r=0$, as in Fig. \ref{fig4}. However, when RR is included in the numerical calculations, the maximum ZeV energies are shown in Fig. \ref{fig9} to be reached for magnetic fields weaker than the on-resonance values when RR is ignored. 

In all cases considered, the reachable maximum ZeV energies in the presence of RR are less than in the absence of RR. Effects of radiation-reaction on the exit kinetic energy are summarized in Table \ref{tab:table1}. The entries are based on that results displayed in Figs. \ref{fig4} and \ref{fig9}. In table \ref{tab:table1}, the deviation is defined by $\Delta K\equiv K -K'$, where $K (K')$ is the kinetic energy without(with) radiation-reaction. Further support for these conclusions is presented in Fig. \ref{fig10}. For every parameter $X\in\{B_s, \omega, \beta_0\}$, Fig. \ref{fig10} shows clearly that the reachable maximum energy is smaller than when RR is neglected, and that the maximum energy is reached, in each case, for a set of RR-based parameters different from the ones related by $r=1$, without RR. The maximum attainable energies of the nuclei of helium are lowered by RR less than both hydrogen and iron nuclei, in agreement with the results shown in Fig. \ref{fig3}. More importantly, the optimal exit kinetic energies can be smaller by much less than an order-of-magnitude, when the radiation-reactions are taken into account, than when resonance is met without radiation-reactions.

Besides the impact, shown in Fig. \ref{fig10}, on them when the RR effects are taken into consideration, the exit kinetic energies are lowered due to deviation from the resonance values of $\omega$ and $\beta_0$. This observation leads to conclusions that agree qualitatively with those based on a spread, $\Delta B_s$, about the corresponding on-resonance value of $B_s$. Results without RR (solid lines) and with RR (dotted lines) are shown together in each respective case, to allow for comparisons to be made. ZeV particles are obtained for values of the parameters $X'\in\{B'_s, \omega', \beta'_0\}$ in windows $\Delta X' \sim 10\%$. For example, scanning the static magnetic field beyond the range between $B'_s-0.118B'_s$ and $B'_s+0.311B'_s$ leads to proton energies up to one order of magnitude lower than the maximum value obtained for $B'_s =0.844B_s$. When RR is taken into account, the maximum kinetic energies are obtained for weaker static magnetic fields, as well as at smaller values of $\omega$ and $\beta_0$. For more accurate comparisons, see Table \ref{tab:table2}, in which differences are shown between the on-resonance values of $X$ (without RR) and the values $X'$ that correspond to maximum energy gain when RR is taken into account. For example, the energy for a proton peaks at a value of $B'_s$ that is about 11.8\% below the corresponding on-resonance value of $B_s$. 

Thus, by including the radiation-reaction effects, the realistic astrophysical conditions have been better simulated in our calculations. It has been demonstrated that, under these conditions, particles in the ZeV range can be generated, even for parameters that violate the resonance condition substantially. 

\section{Discussion}\label{sec:discussion}

The process of single-particle energy gain in CARA will be highlighted further in this section. This will be followed by a discussion of the astrophysical conditions that may lead to ZeV gain.

\subsection{The underlying process of energy gain}

A particle gains zero energy from interaction with an integer number of cycles of a plane-wave radiation field, in the absence of boundaries or a material medium, and under off-resonance conditions, according to the Lawson-Woodward theorem \citep{lw}. Effectively, the particle gains energy during interaction with the positive (accelerating) half of a single cycle, only to lose it entirely during interaction with the following (identical) negative (decelerating) half. In the added presence of a static magnetic field $\bm{B}_s$, however, the Lawson-Woodward theorem gets circumvented and a resonance condition may be met. On resonance, the electric field vector of, say, a circularly-polarized plane wave, gyrates about the direction of propagation at the angular frequency, $\omega$, of the wave. The particle, while in cyclotron motion about the lines of $\bm{B}_s$, senses the radiation field at the Doppler-shifted frequency $\omega_D=\omega\sqrt{(1-\beta_0)/(1+\beta_0)}$. The argument holds just as well for a linearly-polarized radiation field, which may be represented in terms of a superposition of two circularly-polarized ones, of opposite helicity. In either case, auto-resonance guarantees that the vectors $\bm{\beta}$ and $\bm{E}$ in Eq. (\ref{eeq}) maintain the same orientation relative to each other. This means the rate $d{\cal E}/dt$ will remain positive, implying energy gain, during interaction with both halves of every radiation field phase-cycle. Thus, energy of the particle continues to increase monotonically. Barring any unaccounted for perturbations, the resonance condition continues to hold true, due to the constant of the motion expressed by Eq. (\ref{constant}) and the particle continues to be accelerated.

\begin{deluxetable}{rcccc}
\tablenum{2}
\label{tab:table2}
\tablecaption{Off-resonance parameter deviations for moderate energy drop}
\tablewidth{0pt}
\tablehead{
\colhead{} &~& \colhead{H$^{+1}$}  & \colhead{He$^{+2}$} &  \colhead{Fe$^{+26}$}}
\startdata
${B'_s}/{B_s}~(\%)$ &~& 84.8 & 94.4 & 74.8 \\
${\Delta B'_s}/{B'_s}~(\%)$ &~& [-11.8, 31.1] &  [-16.1, 21.2] & [-8.6, 51.9]\\
${\omega'}/{\omega}~(\%)$ &~& 82.0 & 92.4 & 71.2 \\
${\Delta \omega'}/{\omega'}~(\%)$ &~& [-11.7, 29.3] & [-15.6, 22.5] & [$<$-15.7, 46.6] \\
$\gamma^2_0\beta'_0~(\%)$ &~& 83.6 & 95.2 & 71.6 \\
$\gamma^2_0\Delta \beta'_0/\beta'_0~(\%)$ &~& [-12.4, 58.9] & [-16.8, 21.4] & [$<$-16.2, 82.7] \\
\enddata
\tablecomments{Values of the parameters $X'\in\{B'_s, \omega', \beta'_0\}$ for which optimal exit kinetic energies are obtained in the presence of radiation-reaction (read from the data displayed in Fig. \ref{fig10}) compared to the corresponding on-resonance values of the parameters $X\in\{B_s, \omega, \beta_0\}$, calculated from the condition $r=1$ (without radiation-reaction). Shown here also are detuning windows $\Delta X'$ around the values $X'$ in which particle energies are reduced by at most one order of magnitude as compared to the maximum value when the radiation-reactions are included.}
\end{deluxetable}

\subsection{Astrophysical environment for acceleration}

Chances that the parameters $\beta_0$, $\omega$, $B_s$, and $Q/M$, will conspire to satisfy the auto-resonance condition can surely be very small. Nevertheless, this should not be terribly discouraging, in light of the fact that EECR events are quite rare, indeed. For example, the Telescope Array experiment \citep{abbasi2} reported detecting 72 events, only, with energies of 57 EeV or more (1 EeV = $10^{18}$ eV) over the five-year period between 2008 and 2013. It has been estimated that EECRs are detected at a rate of one particle per km$^2$ per century \citep{nagano}. 

This work has not been specific about any known astrophysical environment where conditions for CARA can initially be met. One possibility can be a small region centered on either pole of a compact object, where the right magnetic and radiation fields may be found. To show that the numbers employed in the examples discussed above are not totally improbable, consider for instance that highly powerful GRBs \citep{kumar,wang,guessoum} can have isotropic energies of the order
of $10^{48}-10^{55}$ erg, in addition to being accompanied by the ejection of protons and other particles, pre-accelerated by internal shock waves to relativistic speeds. The radiant energy is mostly beamed into a jet of a few-degree half-angle, and is typically given off over a period of time that can be as small as a fraction of a second to a few seconds (and as large as 17 minutes in rare cases). Thus the radiation-field intensities of $10^{38} - 10^{44}$ W/m$^2$, used in our examples, may be considered reasonable. 

The magnetic field strength in such an environment, and its direction at, and close to, the pole, can be right for the CARA scheme. However, away from the polar caps of a compact object, the magnetic field lines can be extremely curved and the radiation field intensities fall way below what is needed for CARA to work. Typically, $B_s\sim1/z^3$, for $z>$ radius $\sim 10$~km \citep{price,belcz,rosswog,Pulsars}.

During the relatively brief interaction of the particle with the magnetic and radiation fields, Eq. (\ref{Eeq}) implies that, if met initially, the resonance condition continues to hold so long as the conditions outlined in Sec. \ref{sec:theory} are met. Radiation in the astrophysical environments of relevance to this work can be coherent \citep{Huege,kunihito,Gainullin}. On the other hand, a recent study has shown that particles can be efficiently accelerated by incoherent radiation in the wakefield of a plasma \citep{benedetti}.

\section{Conclusions}\label{sec:conc}

Basic elements of the scheme of cyclotron  auto-resonance acceleration (CARA) have been reviewed in this paper and tailored to meet astrophysical conditions of ultra-strong magnetic and super-intense radiation fields, which would work to efficiently accelerate protons and heavier nuclei to ZeV energies. The results include equations giving the trajectory and Lorentz factor of a single particle, in closed analytic form. This has been accomplished in the absence of radiation-reactions. Radiation-reactions have later been approximately included via numerical simulations and their effects on the optimal energy gains have been assessed.

Mega- and giga-tesla magnetic fields, and radiation fields of intensity $10^{38}-10^{44}$ W/m$^2$, have been shown to accelerate nuclei of hydrogen, helium and iron, to ZeV energies, over distances ranging from several hundred meters to many kilometers. Directions of the magnetic field and that of propagation of the radiation have been assumed to be strictly parallel. Magnetic field lines through a small region around the pole (within the polar cap) of a compact object, for example, may be straight and strong enough over a long distance, but can be severely curved elsewhere. The particles have also been assumed to be pre-accelerated inside their source to relativistic speeds before entering the region for further acceleration by CARA. 

Key for the process to work is meeting the auto-resonance condition. It has been shown that, when radiation-reactions are neglected, once the resonance condition is met initially, it would hold almost exactly throughout. In this regime, dependence of resonance upon the initial conditions and magnetic and radiation field parameters has been investigated. It has been shown that the (on-resonance) ZeV energy gained by a particle stays within an order-of-magnitude of that maximum value if resonance is slightly missed due to a spread of a few tens of percentage points in the values of one or more of the parameters $B_s$, $\omega$, or $\beta_0$. It has also been demonstrated via careful numerical calculations that the ZeV energy gains are still maintained, also to within an order-of-magnitude, when radiation-reactions are taken into account. 

\acknowledgments

The authors acknowledge fruitful discussions with Ahmad Hujeirat and Matteo Tamburini. Work of YIS has been supported by an American University of Sharjah Faculty Research Grant (number: AS1811).

\bibliography{Zevatron}
\bibliographystyle{aasjournal}

\end{document}